\documentclass[11pt,twoside]{article}
\usepackage{asp2010}

\resetcounters

\begin{document}

\title{HMI: first results}

\author{Rebecca~Centeno1$^1$, Steve~Tomczyk$^1$, Juan~Manuel~Borrero$^2$, Sebastien Couvidat$^3$, Keiji Hayashi$^3$, Todd Hoeksema$^3$, Yang Liu$^3$ and Jesper Schou$^3$
\affil{$^1$High Altitude Observatory (NCAR), Boulder, CO}
\affil{$^2$Kiepenheuer-Institut  f\"ur Sonnenphysik, Freiburg, Germany}
\affil{$^3$ Stanford University, Stanford, CA}}

\begin{abstract}

The Helioseismic and Magnetic Imager (HMI) has just started producing data that will help determine what the sources and mechanisms of variability in the Sun's interior are. The instrument measures the Doppler shift and the polarization of the Fe I 6173 \AA\ line, on the entire solar disk at a relatively-high cadence, in order to study the oscillations and the evolution of the full vector magnetic field of the solar Photosphere. 

After the data are properly calibrated, they are given to a Milne-Eddington inversion code \citep[VFISV,][]{juanma_vfisv} whose purpose is to infer certain aspects of the physical conditions in the Sun's Photosphere, such as the full 3-D topology of the magnetic field and the line-of-sight velocity at the solar surface.
We will briefly describe the characteristics of the inversion code, its advantages and limitations --both in the context of the model atmosphere and the actual nature of the data--, and other aspects of its performance on such a remarkable data load. Also, a cross-comparison with near-simultaneous maps from the Spectro-Polarimeter (SP) onboard Hinode will be made. 

\end{abstract}

\section{Introduction}

The Solar Dynamics Observatory (SDO) is NASA's first mission of the Living with a Star program, which is designed to study the causes of solar variability and its impacts on life and humanity's technological development. 
Solar variability is intimately related to magnetic activity, so the main goals of the program  are to understand the mechanisms that produce these fields and drive them to the surface, and be able to predict when and where the energy stored in them is eventually going to be released in the form of particle ejections and changes in the solar irradiance.

\noindent SDO was launched from Cape Canaveral on February 11, 2010, carrying three instruments on board:
the Atmospheric Imaging Assembly (AIA), the Extreme ultraViolet Experiment (EVE) and the Helioseismic and Magnetic Imager (HMI).
The nominal lifetime of the mission is just over 5 years, with an extension of up to 10 years.
The spacecraft follows a geosynchronous orbit (24h period at 36000 km) passing over over the ground station in White Sands Missile Range in New Mexico once a day, to where it downloads the more than 1 TB of data a day that the three instruments produce altogether.

\section{Helioseismic and Magnetic Imager}

\begin{figure}[!t]
\begin{center}
\includegraphics[scale=0.37]{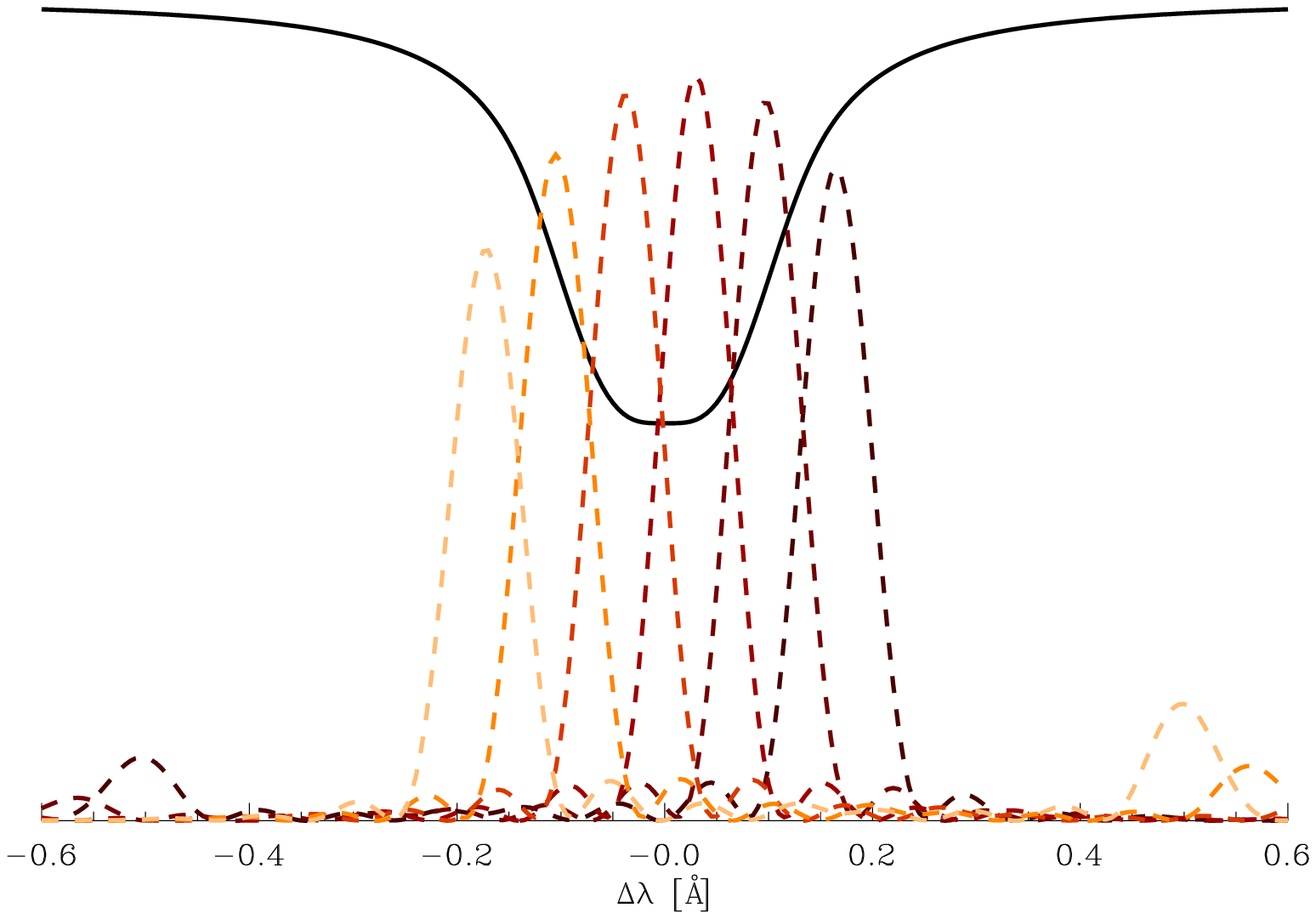}
\includegraphics[scale=0.37]{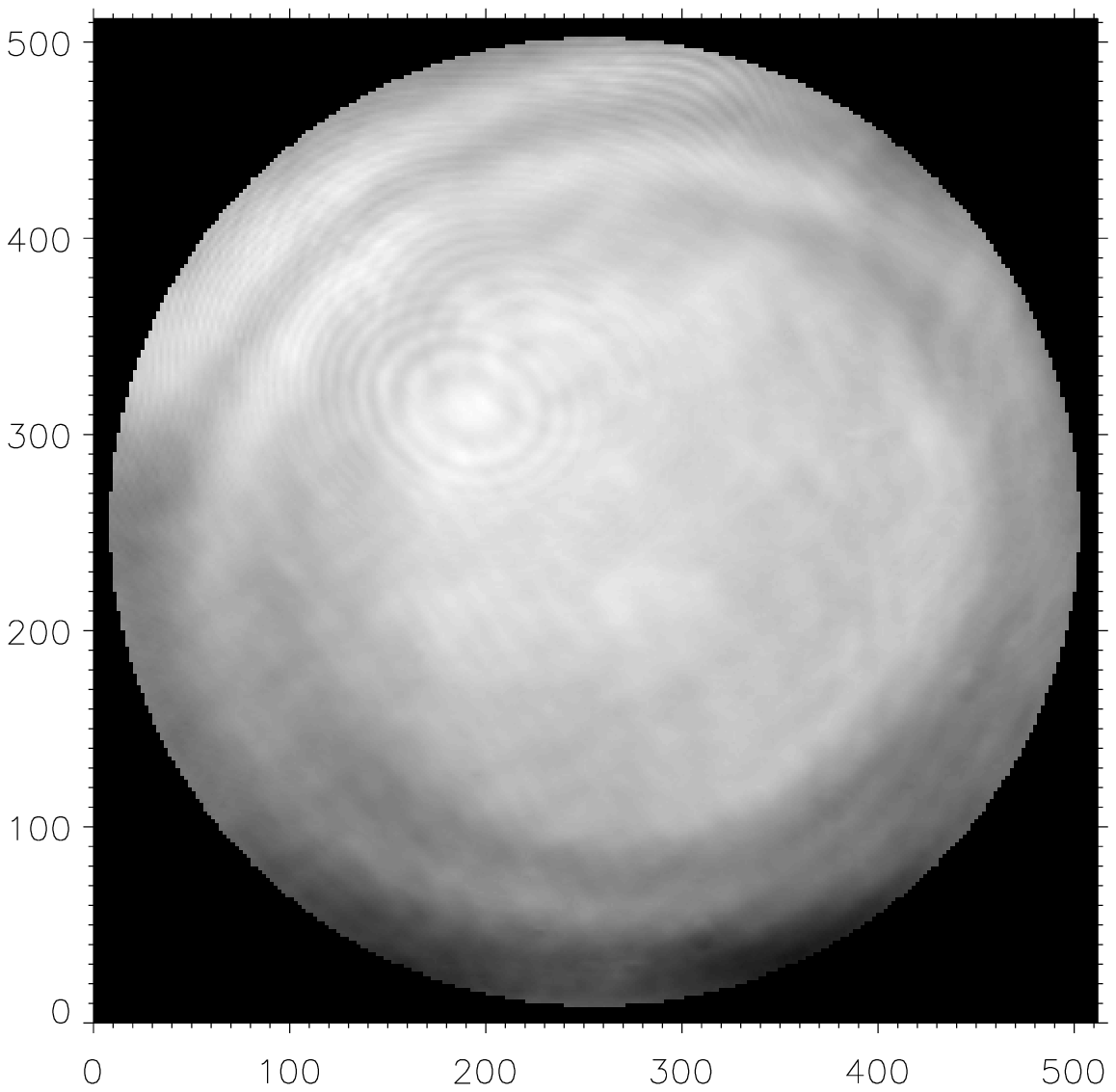}
\caption{Left: representation of HMI's filter profiles (dashed) and their position relative to the Fe {\sc i} 6173 \AA\ spectral line. Right: Changes in the position of the filter profiles across the field of view. The grayscale represents inferred Dopplershifts values from the aparent change in position of the spectral line. The scale goes from $0$ to $700$ m/s}
\label{fig:filters}
\end{center}
\end{figure}

HMI was designed to study the oscillations and magnetic fields at the photosphere of the Sun. It is not only a successor to the Michelson-Doppler Interferometer (MDI, onboard SOHO), but also an upgrade, since HMI is able to observe the full Stokes vector of a magnetically sensitive spectral line on the whole disk of the Sun with a 1 arcsec spatial resolution and a 90 second cadence.

HMI is a filter instrument, consisting of a refracting telescope, a polarization selector, an image stabilization system, a narrow-band tunable filter and two $4096\times4096$ pixel CCD cameras, one devoted to the full Stokes vector and the other one to Stokes I/V measurements only. Images are made in a sequence of wavelength tuning and polarization selection at a 4-second cadence for each camera. 

\noindent The instrument operates by scanning through 6 wavelength positions along the spectral line. The need to measure the Doppler velocity and magnetic field with very limited spectral information sets strict requirements on the choice of the spectral line. 
SDO has a dynamic range of 6.5 km/s, so this imposes that there is a clean continuum around the measured line. Also, Helioseismology requires deep spectral features so that the Doppler sensitivity is large, while the Zeeman diagnostics of magnetic fields call for large Land\'e factors to ensure high magnetic sensitivity. For all these reasons, the chosen spectral line for HMI was the Fe I 6173 \AA\ line.

The filter system consists of an entrance window, a blocking filter, 5 Lyot elements (one is tunable) and 2 Michelson interferometers. Together, the elements of the system enable narrow-band filtergrams to be made across the 6173 \AA\ line by co-tuning one Lyot element and the two Michelsons. The final filter width is 76 m\AA\ and the filter range is 690 m\AA. On the left side of Fig. \ref{fig:filters}, the black line represents the spectral feature while the dashed coloured lines show the filter profiles of the instrument. HMI integrates the light under each of these filter profiles creating 1 full disk filtergram for each of the 6 wavelengths.
However, the filter profiles are not identical for all pixels of the field of view (FOV).
The right panel of Fig. \ref{fig:filters} shows the FOV of the HMI instrument. The shades of gray represent the wavelength position of one filter profile as it changes across the FOV. These variations will translate into a sampling of the spectral profile that occurs at slightly different wavelengths for each pixel on the image, which is only an instrumental artifact but will be interpreted by any spectral line inversion code as changes in the LOS velocity across the field of view (when working under the assumption that the filter profiles are identical for all pixels). The scale in the image corresponds to Doppler shifts between 0 and 700 m s$^{-1}$.

Changes in amplitude, shifts and shape of the HMI filter profiles from pixel to pixel require a careful calibration of the instrument for proper interpretation of the data.

\subsection{Data pipeline}

\begin{figure}[!t]
\begin{center}
\includegraphics[scale=0.32]{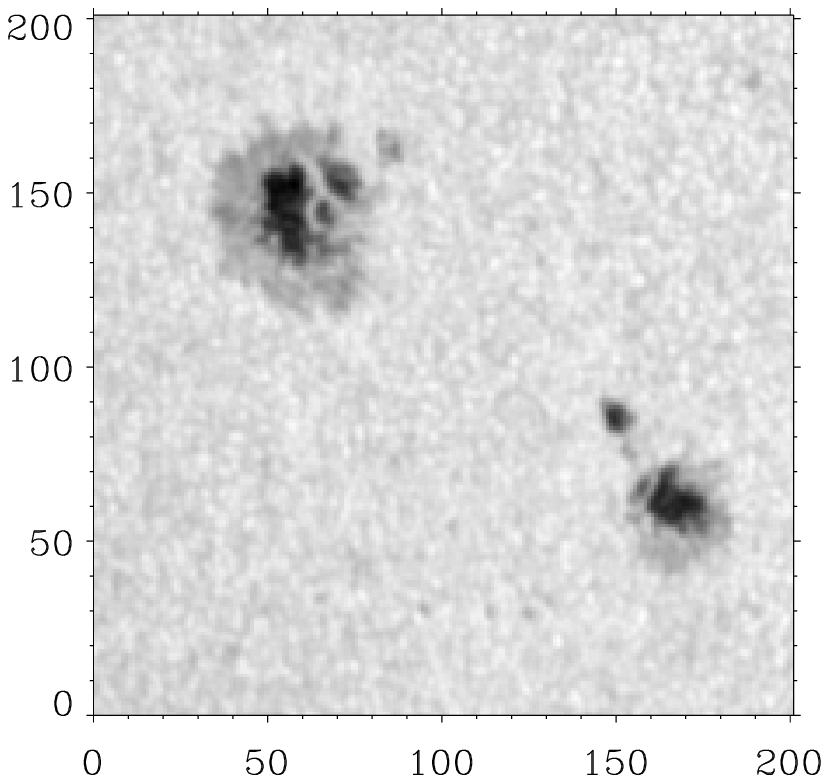}
\includegraphics[scale=0.32]{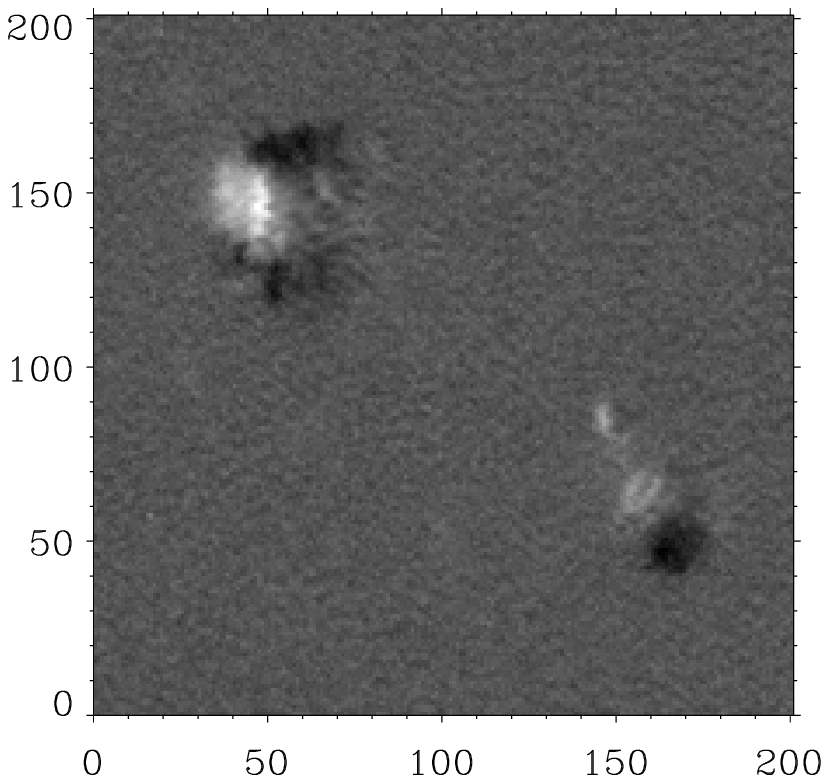}
\includegraphics[scale=0.32]{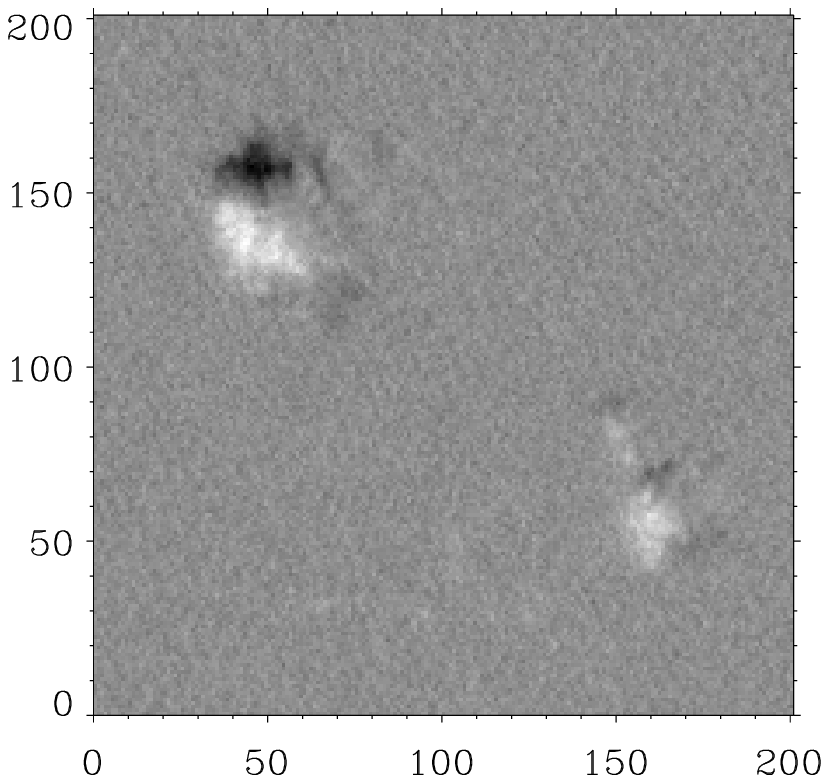}
\includegraphics[scale=0.32]{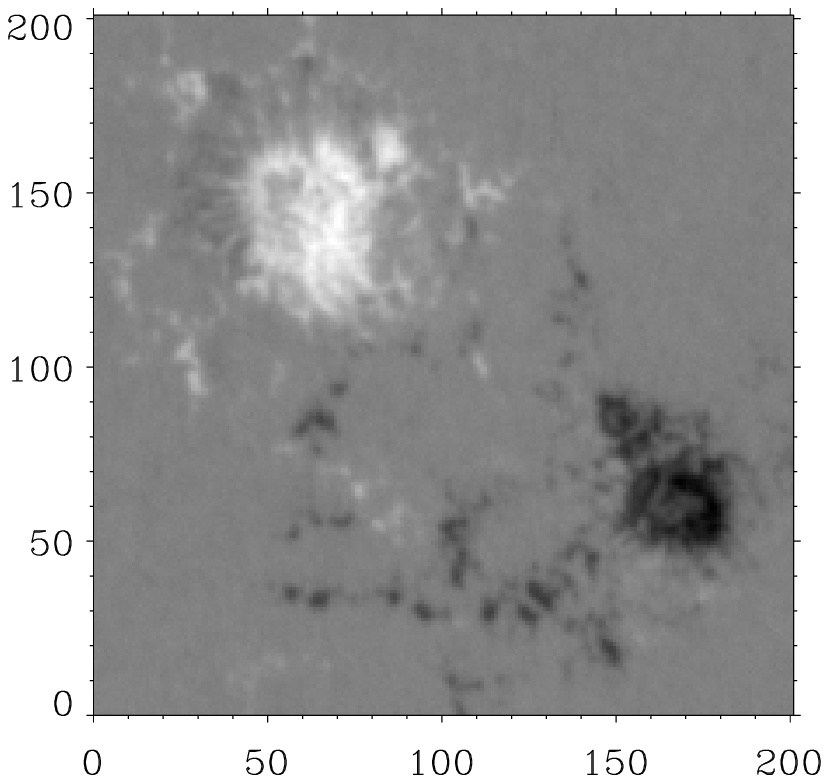}
\caption{Images of Stokes I, Q, U and V for a $100\times100$ arcsec$^2$ region containing a pair of sunspots. The different Stokes parameters are shown for different filter wavelengths.}
\label{fig:stokes}
\end{center}
\end{figure}

The data that arrive at the ground station in White Sands are subject to several levels of processing before being made available to the public.
Level 0 data consist of a series of filtergrams taken sequentially within 90 - 120 seconds. For each of the 6 wavelength tuning positions, 4 to 6 images in different polarization modulation states are taken, depending on the chosen modulation scheme. 

\noindent When Level 0 data are received from the ground station in NM they still contain overscan rows and columns that need to be removed. Certain header values have to be measured and set (pixel scales, image center, etc) and images are then corrected for dark current, flatfield, exposure time and cosmic ray hits.
After that, they are demodulated to obtain the Stokes profiles, and cross-talk correction and temporal averaging are applied to produce Level 1.5 data.

Level 1.5 data consist of a series of images of the full disk of the Sun in different wavelengths and polarizaton states. Fig. \ref{fig:stokes} shows Stokes I, Q, U and V for a portion of the Sun containing an active region (the different Stokes parameters are shown for different wavelength positions of the filter profiles).

For each of the Stokes parameters, HMI scans across the 6173 \AA\ line producing 6 filtergrams with a 75 m\AA\ width and a 70 m\AA\ separation. For each pixel of ($0.5\times0.5$ arcsec$^2$) the full Stokes vector, with 6 wavelength points each, is available.
From this information we need to retrieve Doppler velocities and the full vector magnetic field for every position on the solar disk.
The Level 1.5 data are given to a spectral line inversion module (Very Fast Inversion of the Stokes Vector, VFISV) in the pipeline, which interprets the observed Stokes profiles and produces the physical parameters of the atmosphere in which they were generated, according to a Milne-Eddington model.
Then, the vector magnetic field is desambiguated and a series of data products are produced and made available to the scientific community through Stanford’s JSOC database.

\section{The inversion module}

\begin{figure}[!t]
\begin{center}
\includegraphics[scale=0.32]{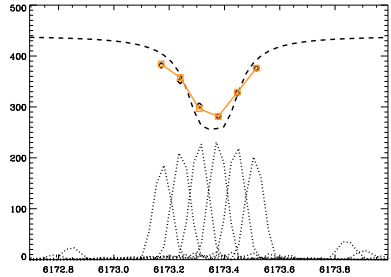}
\caption{Synthesis of HMI spectral line. The black squares represent the observed intensity spectral profile for a given pixel. The code first synthesizes a high spectral resolution line  (dashed) that is then filtered with the corresponding HMI filter profiles (dotted) producing the synthetic HMI-like intensity profile (orange diamonds and line).}
\label{fig:inversion}
\end{center}
\end{figure}

The spectral line inversion code \citep[VFISV,]{juanma_vfisv} is based on a Milne-Eddington solution to the polarized radiative transfer equation, which assumes constant properties of the atmosphere with height, except for the source function, which is allowed to very linearly with optical depth. The generation of polarized radiation is understood within the Zeeman effect  scenario.
This simple model is parametrized with 10 physical parameters, two of which are kept constant in the HMI pipeline inversion module (the damping and the magnetic filling factor).
The inversion code works on a Levenberg-Marquardt scheme. Given an initial guess model atmosphere, the code iteratively modifies it until it generates a set of Stokes profiles that best fits the observed data in a least squares sense. 
However, since the HMI instrument integrates the light under a set of six filter profiles -- and these vary across the FOV--, the inversion code has to take them into account.
The way to do this (schematized in Fig. \ref{fig:inversion}) is to synthesize a spectral line with a relatively high spectral resolution. Then, the HMI filter profiles are applied to the synthetic spectral line to generate HMI-like data (six spectral positions for each of the 4 Stokes profiles), which are compared to the observed data in the iterative process.

\noindent The magnetic field strength, orientation and azimuth, together with the magnetic filling factor are then passed on further down the pipeline to the disambiguation module, which produces the best, continuous solution for the vector field on certain regions of interest of the solar disk (active regions).

\section{Comparison with Hinode SP}

Filter instruments are usually optimized to provide high spatial resolution and high cadence data, which are very useful when studying certain aspects of small scale  and/or highly dynamic features in the solar atmosphere. Spectrograph-based instruments are better at resolving spectral details and having higher spectral sensitivity, but do not provide the possibility of aquiring a two-dimensional instantaneous image of the Sun, let alone at a high cadence. There is always a compromise that one has to make among the desirable features of an instrument -- and there are many to choose from: high spatial resolution, high cadence, large polarimetric sensitivity, 2D large field of view, high spectral resolution, multiple wavelength bands, etc.

To place HMI data into perspective, we decided to make a rough comparison with one of the state-of-the-art reference instruments in Solar Physics of the past several years: the Spectro-Polarimeter onboard Hinode (SP).
\noindent When comparing SP and HMI one has to be aware of the differences in operation mechanism between both instruments. While HMI makes a series of filtergrams in 4 polarization states and 6 different wavelengths across the spectral line, SP constructs a bi-dimensional image of the solar surface by scanning across it step by step (however, the spectral and polarimetric information for the pixels along the spectrograph slit can be considered simultaneous). 
\noindent Thus, for SP to construct a 150 arcsec image in its fast operation mode, it takes over half an hour, while for HMI to retrieve the full Stokes vector for the entire solar disk, it takes 90 seconds.

For the purpose of the comparison, we took an SP map from April 6 2010 that contained an active region, and chose the corresponding HMI data-set that was closest in time to the scan at the center of the SP map. The properties and major differences between both datasets are summarized in Table \ref{tab:instruments}.

\begin{table}
\begin{center}
\begin{tabular}{c|cc}
Features & SP & HMI\\
\hline
Spatial resolution (arcsec)  & 0.3 & 1\\  
Spatial sampling (arcsec) & 0.3 & 0.5 \\  
FOV (arcsec$^2$) &  $150\times150$ & Full disk \\
Time per map (s) & 2500 & 90 (720 averaged) \\
Spectral coverage & Fe {\sc i} 6301.5, 6302.5 \AA & Fe {\sc i} 6173 \AA  \\
Spectral sampling  (m\AA)  & 21.5 & $\sim 70$\\
Inversion code & MERLIN & VFISV\\
\end{tabular}
\end{center}
\caption{Differences in the main properties of the HMI and SP datasets of Apr 6, 2010.\label{tab:instruments}}
\end{table}

The spatial resolution of both instruments is significantly different (there is over a factor 3 difference between them, although Hinode's fast mode under-samples its nominal resolution). The time that SP takes to complete a 150x150 arcsec$^2$ map in fast mode is around 35 minutes, while HMI produces a full disk, full Stokes dataset in 90 seconds (although we use 12-minute averages).

Due to the very different nature of the data from both instruments, the comparison was made over the atmospheric parameters inferred from the full Stokes inversion of the data-sets, rather than over the spectral profiles themselves. The standard inversion codes for both instruments are based on a Milne-Eddington model atmosphere, thus including the same approach and physical assumptions. However, they are different codes written in different languages with subtle modifications and optimizations for the specific datasets they are written for. SP inversions were carried out with the MERLIN code (http://www.csac.hao.ucar.edu/csac/nextGeneration.jsp\#merlin) while HMI data were inverted with the VFISV module.

All the data and inverted maps from SP were obtained from the HAO/CSAC website\footnote{http://www.csac.hao.ucar.edu/csac/dataHostSearch.jsp}.

\subsection{Alignment}

\begin{figure}[!ht]
\begin{center}
\includegraphics[scale=0.35]{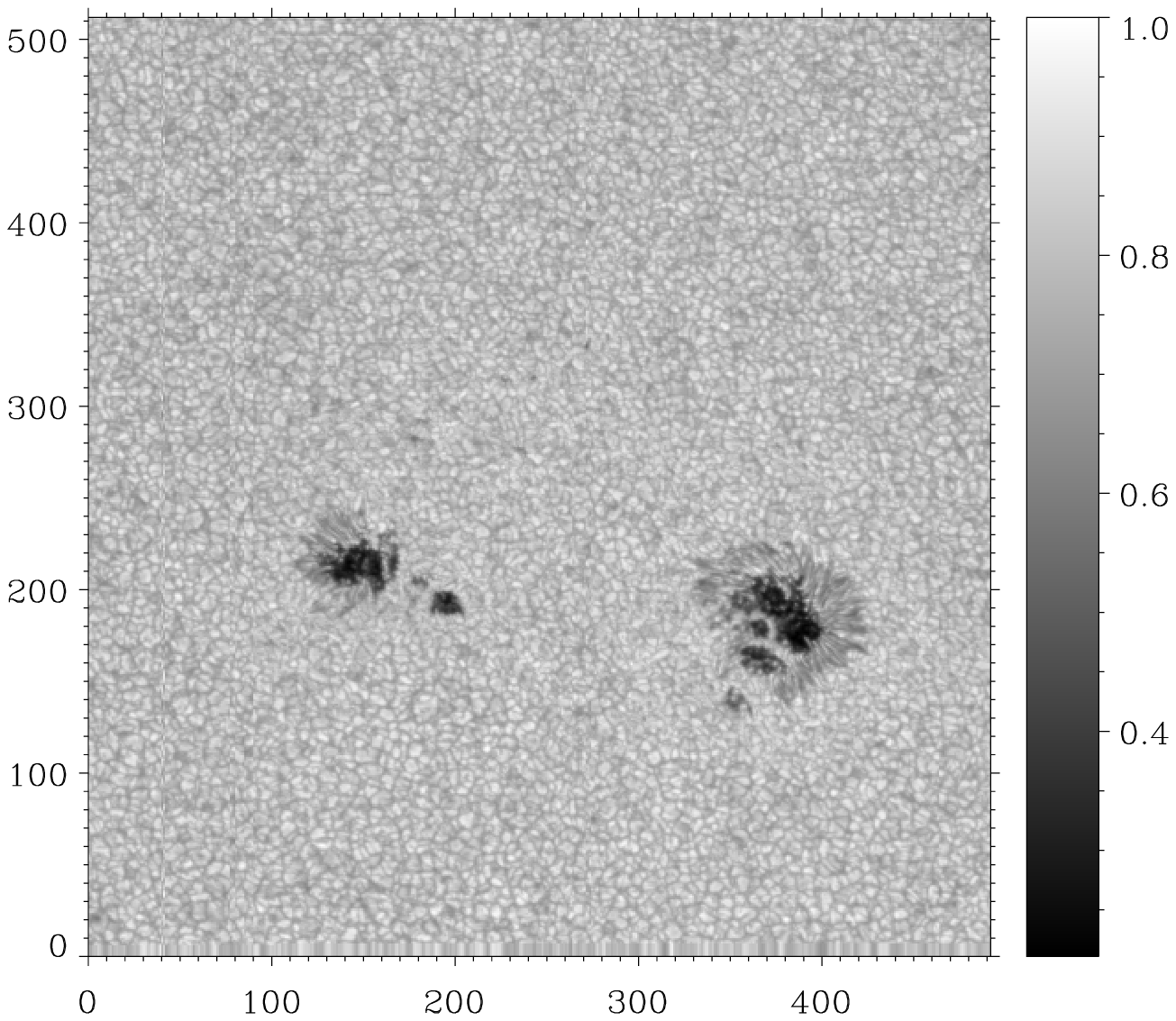}
\includegraphics[scale=0.35]{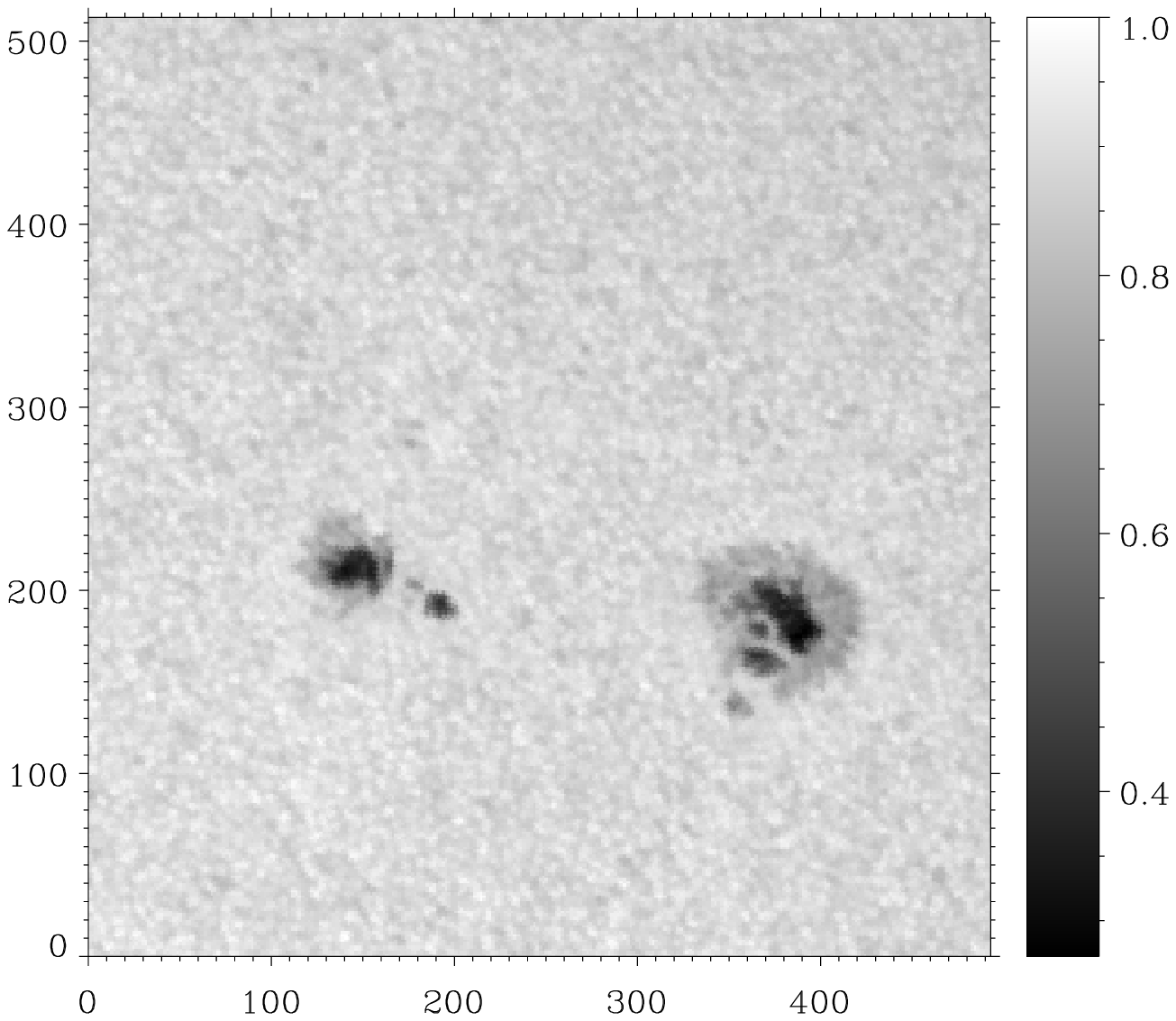}
\includegraphics[scale=0.35]{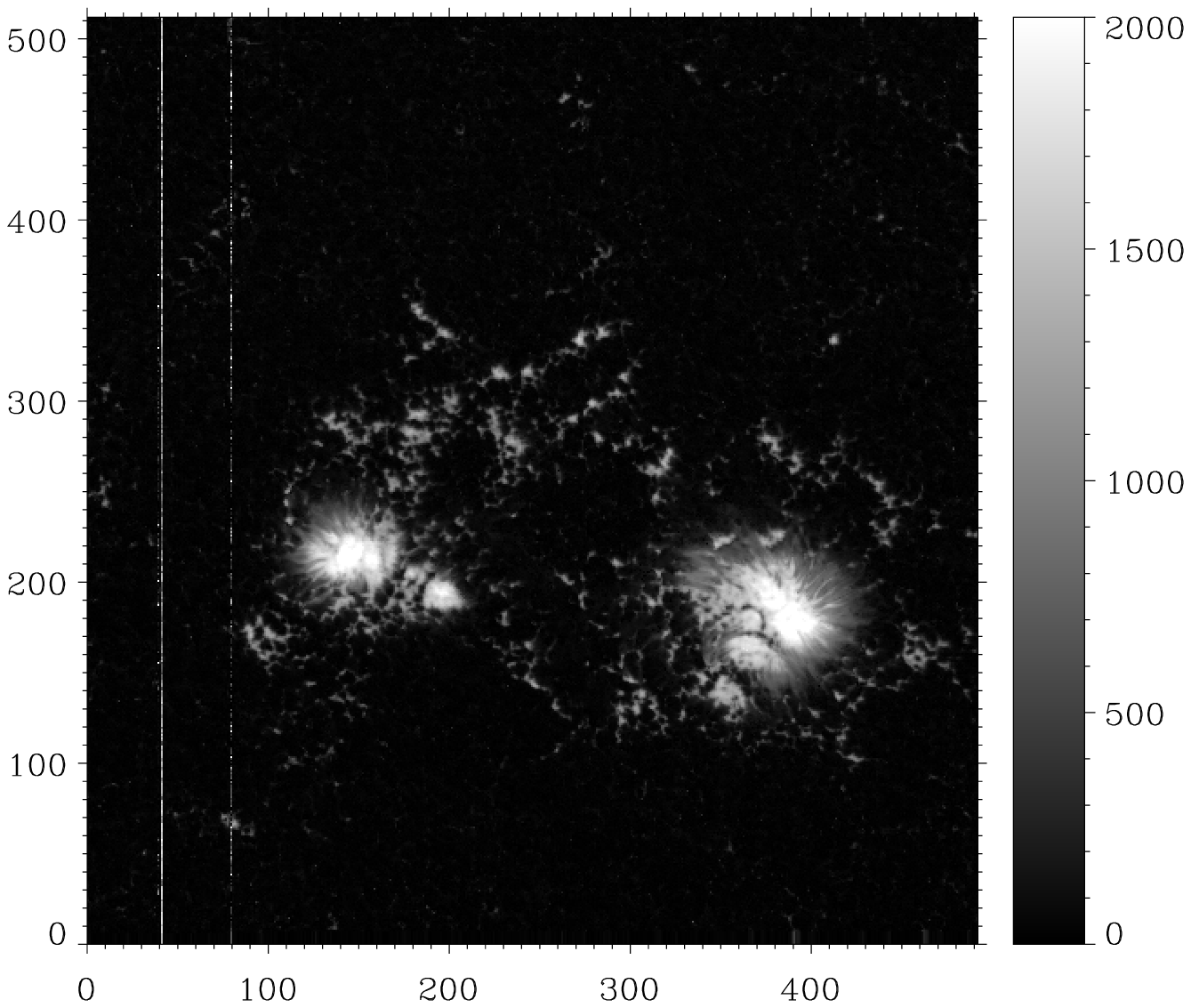}
\includegraphics[scale=0.35]{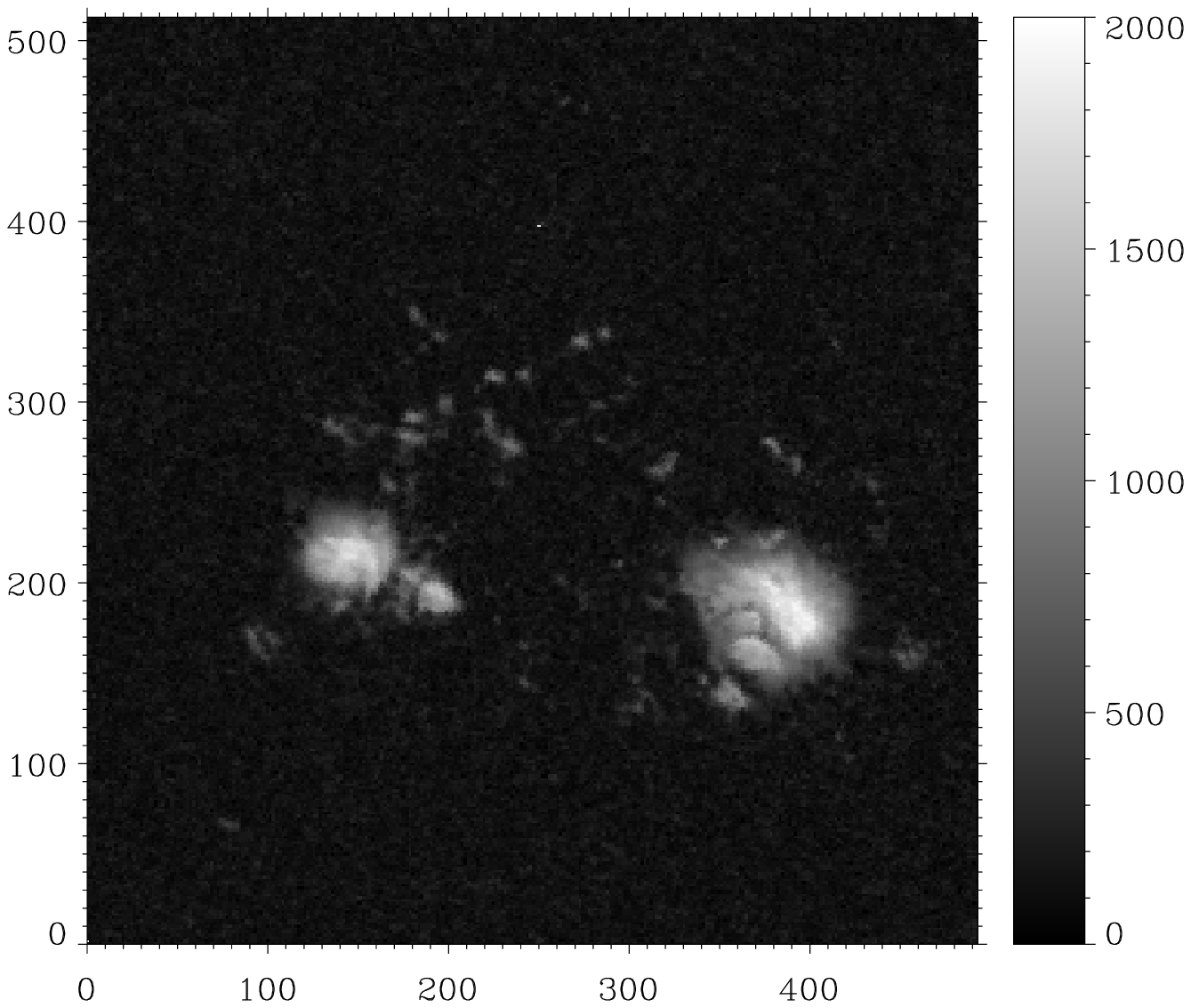}
\includegraphics[scale=0.35]{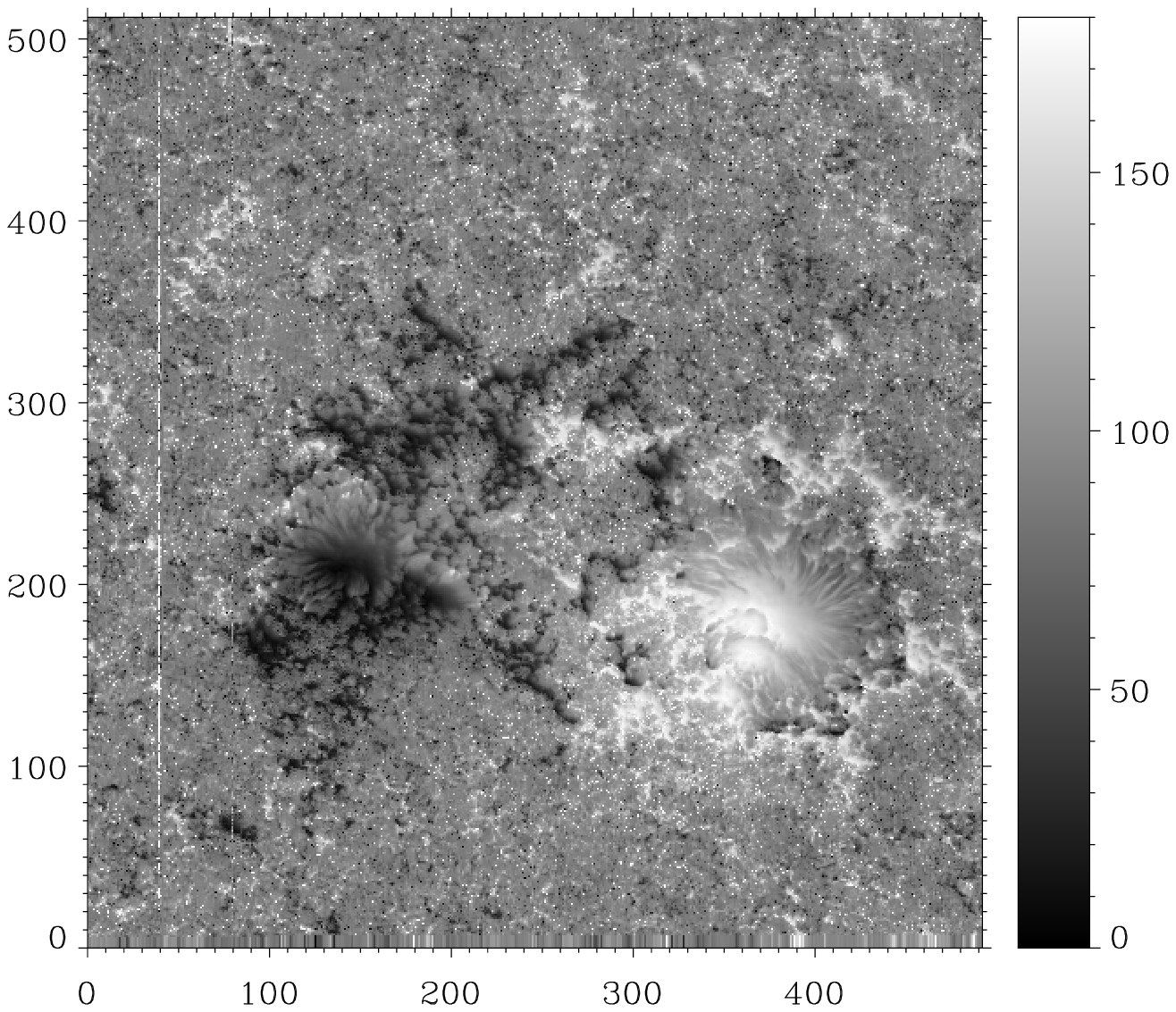}
\includegraphics[scale=0.35]{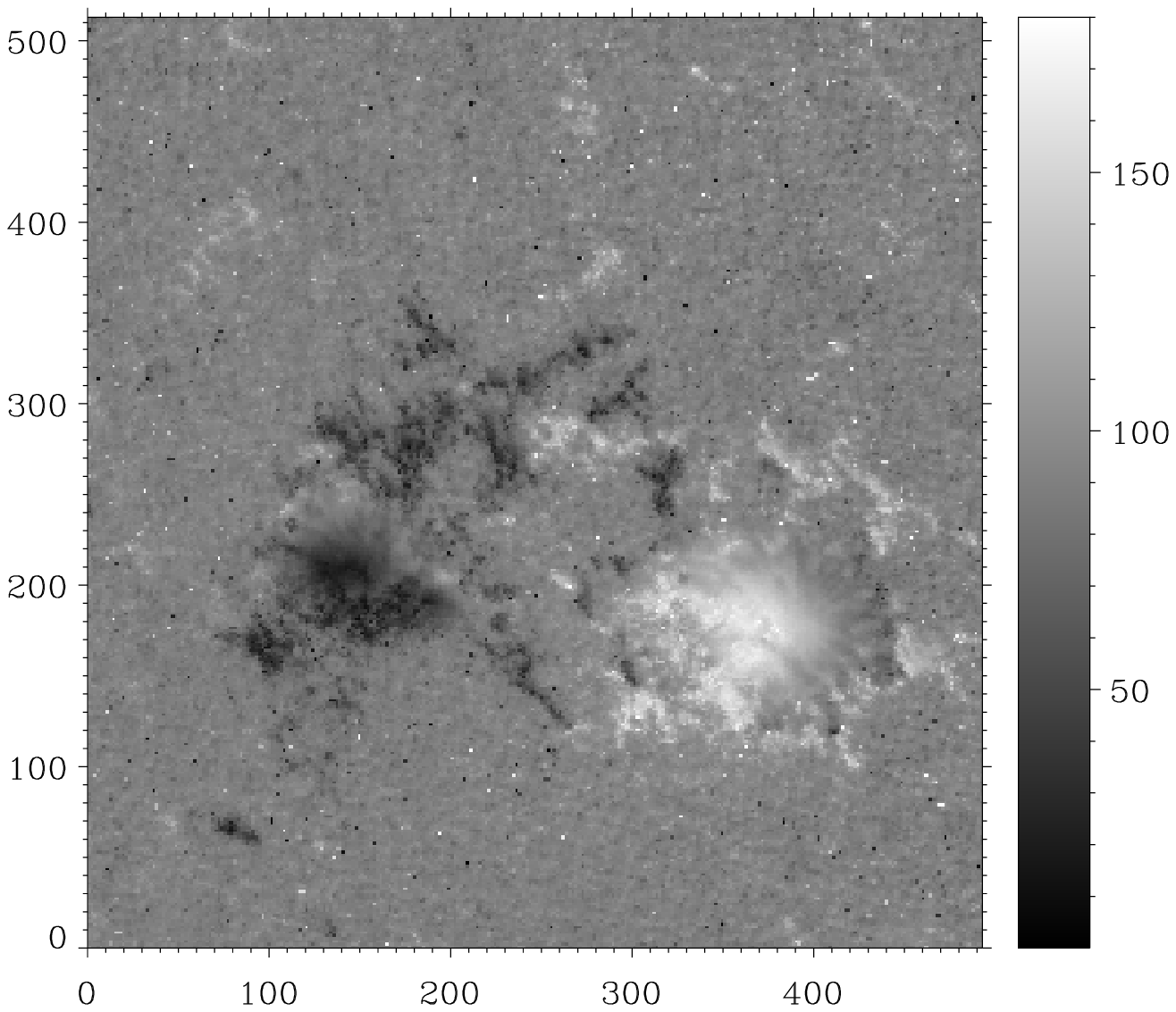}
\includegraphics[scale=0.35]{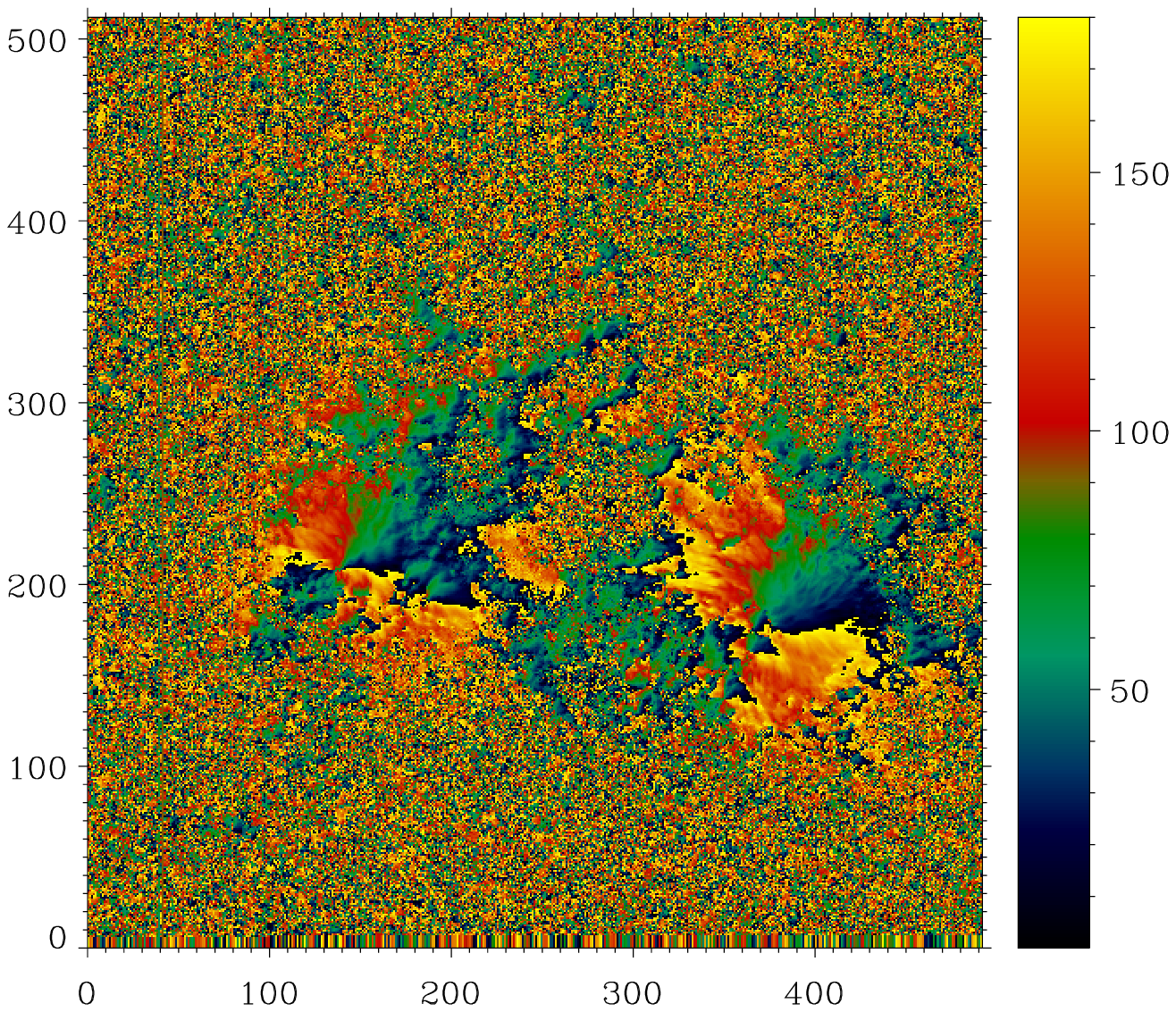}
\includegraphics[scale=0.35]{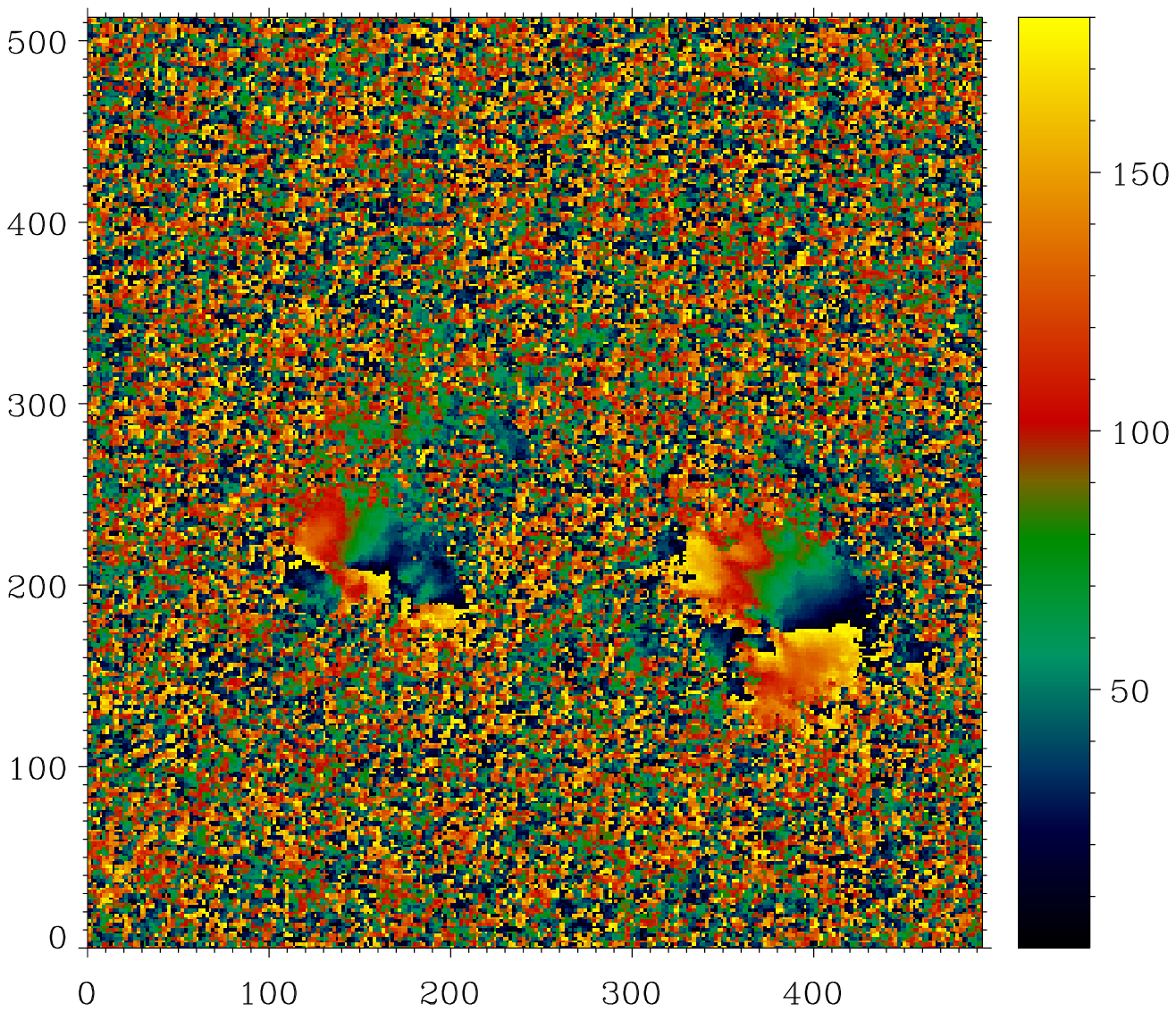}
\caption{Comparison between SP and HMI. Rows 1 through 4 show images of the continuum intensity, magnetic flux density (G), and magnetic field inclination (degrees) and azimuth (degrees). The left column corresponds to SP and the column on the right to HMI.}
\label{fig:comparison}
\end{center}
\end{figure}

The first step in any data-set comparison is the alignment of the data. Due to the different nature of the instruments (the HMI data-set is 12-minute averaged set of filtergrams, and the SP map scanned the 150 \arcsec of the FOV in $\sim$35 minutes), simultaneity is never achieved. The evolution of granulation happens at shorter time-scales than these, therefore granulation cannot be used to cross-correlate the images. Sunspots and pores, on the other hand, are relatively long-lasting features, so continuum images can be used to match and align the areas of the map where these objects exist. However, for the remaining area of the field of view, faculae and network will stand out in Stokes V images, rather than in the continuum intensity. Hence, for comparison purposes we used the SP continuum and circular polarization maps (produced by the data-processing package ${\rm sp\_prep}$) and the I0 and V2 filtergrams from the HMI data-set (where I and V refer to the Stokes profiles, and the numbers refer to the filter profile, ranging from 0 to 5 in decreasing wavelength order across the spectral line).

The FOV of the SP map is $150\times150$ arcsec that contains an active region with several sunspots. After selecting and cutting out the region of interest from the HMI map, we applied a rotation and a pixel scaling to roughly align the features by eye. 
This is only a rough alignment. When the images are superimposed, small shifts between certain areas of the images are evident. The SP image was not corrected for the drifting and breathing of the slit-scan mechanism \citep{centeno_hinode}, resulting in differential misalignments across the FOV.

\subsection{Comparison of inversion results}

\begin{figure}[!ht]
\begin{center}
\includegraphics[scale=0.35]{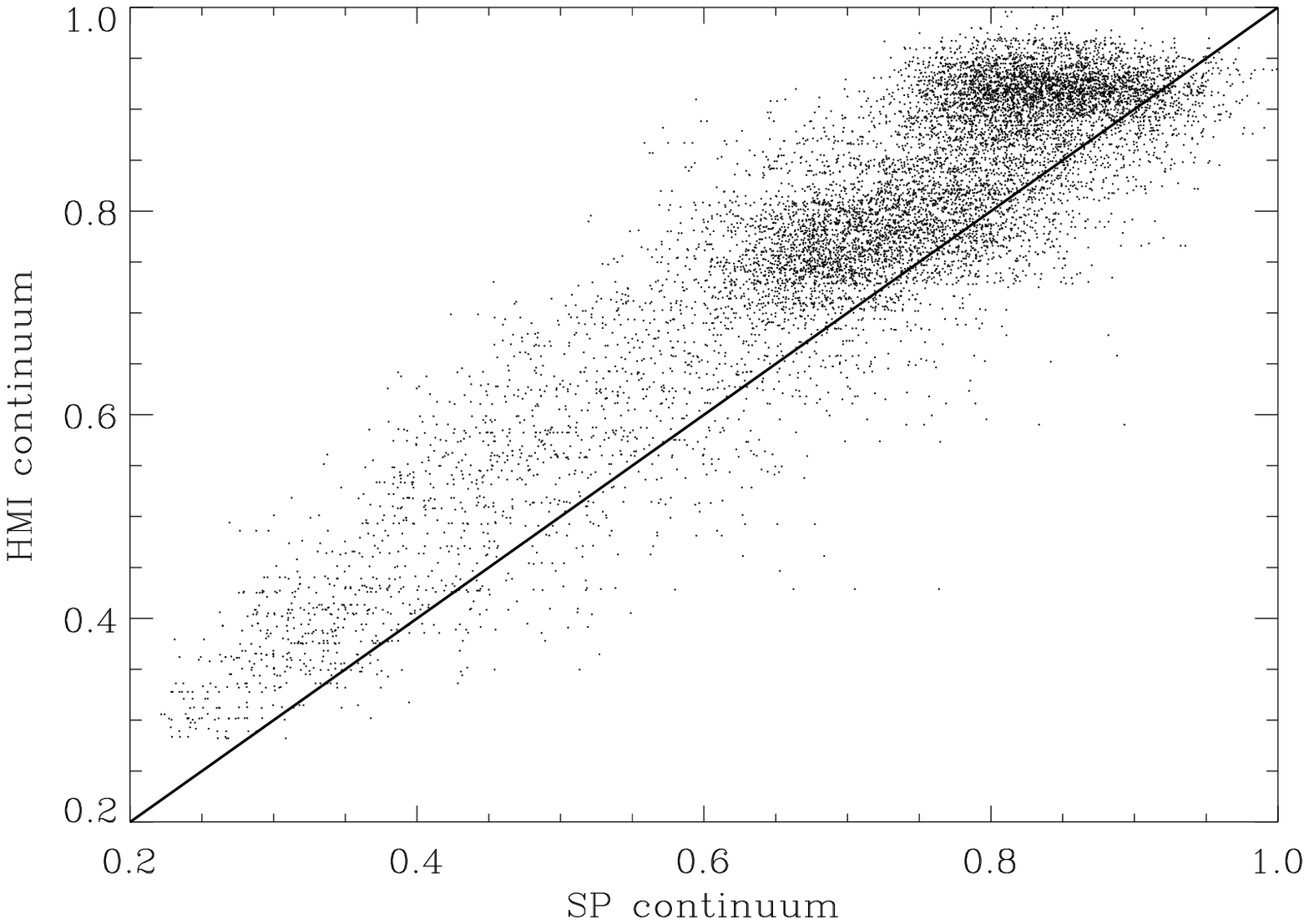}
\includegraphics[scale=0.35]{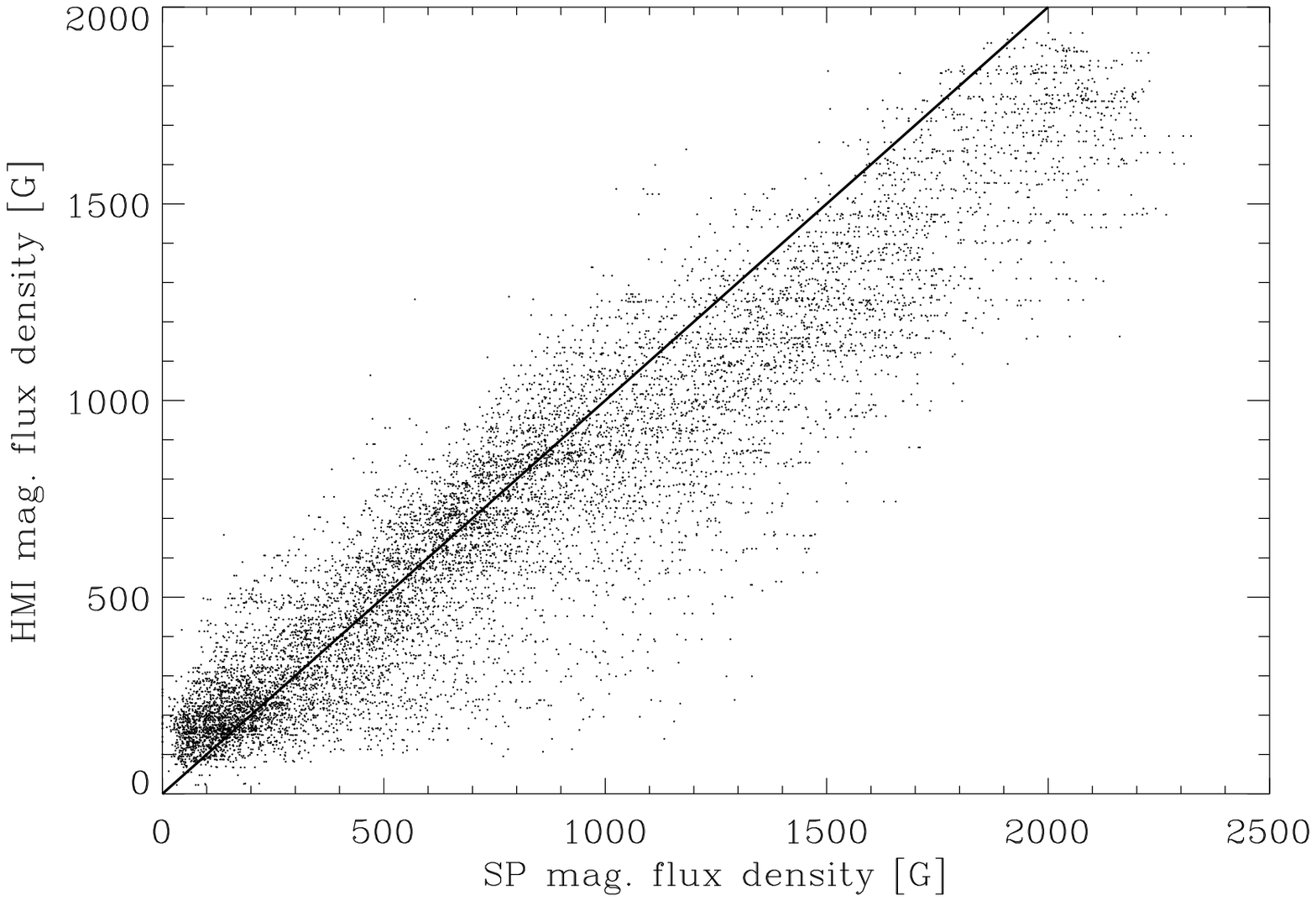}
\includegraphics[scale=0.35]{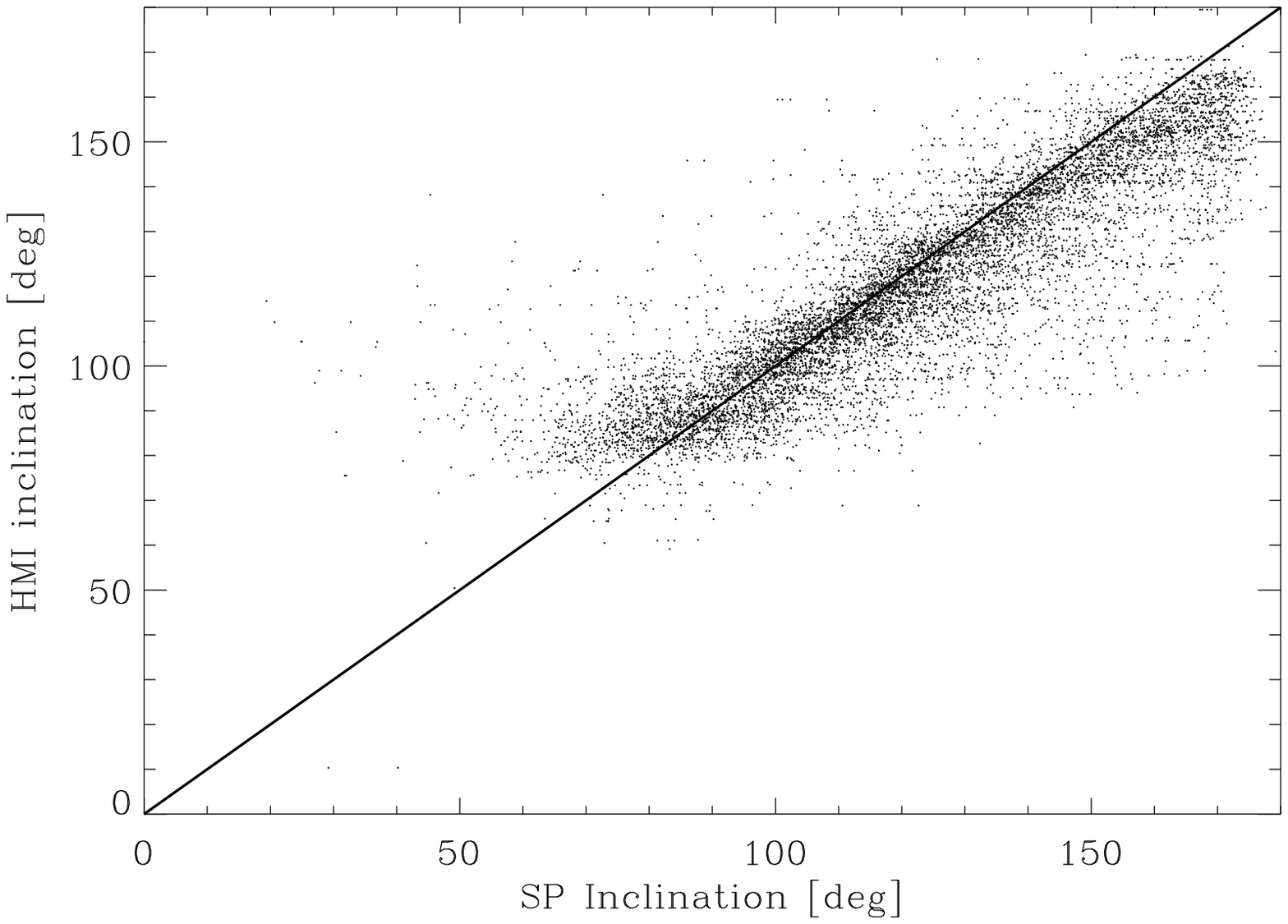}
\includegraphics[scale=0.35]{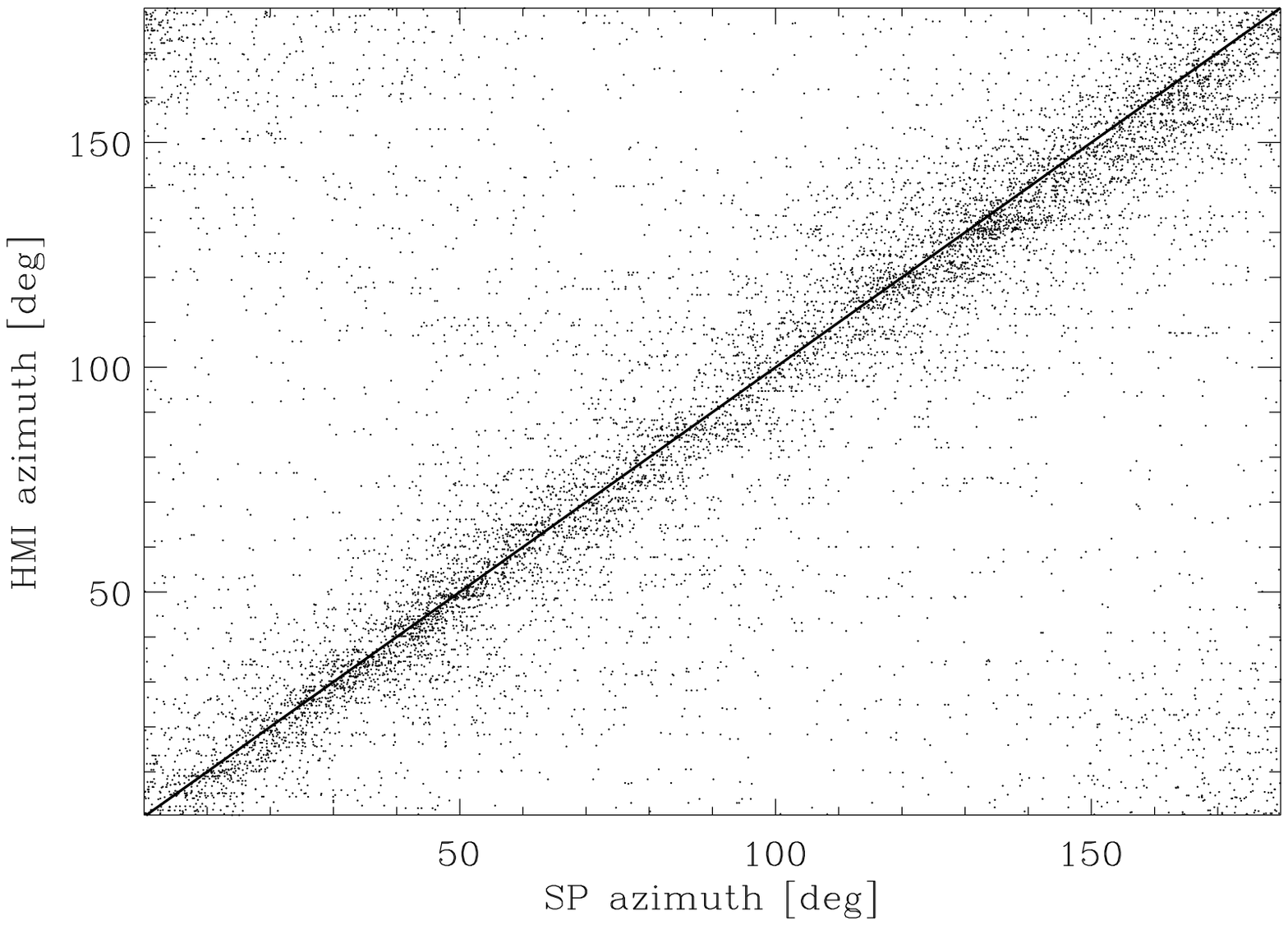}
\caption{Scatter plots. From left to right and top to bottom, scatter plots of the continuum intensity, magnetic flux density, magnetic field inclination and azimuth. X-axis corresponds to SP and y-axis to HMI. The solid line represents the 1:1 correspondence.}
\label{fig:scatter}
\end{center}
\end{figure}

In order to compare the magnetic properties inferred from the full Stokes inversion of both datasets, we directly applied the co-alignment parameters derived from the previous procedure to the inversion results, i.e., the maps of magnetic flux density (magnetic field strength times filling factor), magnetic field inclination and magnetic field azimuth. Figure \ref{fig:comparison} shows the results of the inversions for SP (left) and HMI (right). The top row is the continuum intensity used to calibrate the alignment, while rows 2 through 4 show the magnetic properties (flux density, inclination with respect to LOS, and azimuth).
In the continuum image, most of the mismatch between the two instruments is due to the difference in spatial resolution. However, small misalignments and the lack of simultaneity also play an important role.
Scatter plots in Figure \ref{fig:scatter} show the differences in a more quantitative manner. A square sub-field around the big sunspot was selected to construct them. The top left panel shows the scatter plot of the continuum intensity (with SP on the x-axis and HMI on the y-axis) normalized to the maximum intensity in the sub-field. The solid line represents the 1:1 correspondence, and any desviation from that is due to misalignments, non-simultaneity, differences in spatial and spectral resolution, different sensitivity of the spectral lines to temperature and various other instrumental effects. Also, the HMI I0 filtergram does not correspond to the continuum -- it integrates the light in the far blue wing of the spectral line (see Figure \ref{fig:filters} left). However, the SP continuum does correspond to a region of the spectrum where there is no spectral feature.

The second row in Figure \ref{fig:comparison} shows the magnetic flux density ($\alpha\, B$, where $\alpha$ is the filling factor and $B$ the magnetic field strength) for SP (left) and HMI (right). HMI inversions are carried out assuming a constant filling factor of 1, so by default, the inversion code retrieves the flux density, rather than the field strength. 
There is a very good correspondence between the two images, the main differences being that SP measures larger magnetic flux densities and HMI has a larger noise level. Looking at the scatter plot in the top right panel of Fig. \ref{fig:scatter}, it is obvious that the noise level for HMI is around 100G while it is of the order 20G for SP. HMI shows systematic lower $\alpha \, B$ values than SP in the kG regime. This is partly due to the lower spatial resolution of the HMI instrument, that leads to more polarization signals cancelling out in the same pixel. However, it is yet to be determined how much of this is due to Zeeman cancellation of signals and how much of it has other explanations.

The third and fourth rows of Fig. \ref{fig:comparison} show the magnetic field inclination and azimuth retrieved from both instruments. The bottom row of Fig. \ref{fig:scatter} shows the corresponding scatter plots for these pairs of maps (inclination on the left and azimuth on the right) in the selected subfield around the big sunspot.
Since we are analyzing one of the polarities of the active region, the inclination values go from 90 to 180 degrees only. The agreement is striking and the deviations are most probably due to the differences in spatial and spectral resolution and polarimetric sensitivity. There is a slight deviation of HMI towards less vertical fields when approaching 180 degrees.

\noindent The noise in the linear polarization (Stokes Q and U) will be interpreted by the spectral line inversion code as transverse fields of several dozens of gauss \citep{juanma_kobel}. However, similar noise in Stokes V will lead to vertical fields of only a few gauss. So the noise has two effects: it tends to tilt fields horizontally and it will produce a ground level field different from zero (even if the polarization signals are only due to noise). This explains, partly, why the noise level for the flux density retrieved from HMI data is so large. It could also be responsible for the deviation in inclination when approaching vertical fields.

\noindent The results for the azimuth are also considerabily consistent given the caveats already mentioned. With a certain amount of scatter, the dots are clearly concentrated along the solid line. The random spread-out scatter points mostly come from granulation area outside the strong fields of the sunspot, or wherever the field is primarily vertical and there is no information in the linear polarization signals to tell the inversion code where the azimuth lays. The dots concentrated in the corners diganonally opposite to the 1:1 line, are due to the 180 degree azimuth ambiguity (which has not been accounted for in this analysis).

\section{Conclusions}

To summarize the reasons for the discrepancies it is necessary to understand the different nature of the data provided by both instruments. To start with, HMI and SP observe different spectral lines, and the sensitivity of these to the physical and thermodynamical properties of the atmosphere are not identical. The operation mechanisms of the instruments are such that simultaneity is never properly achieved, and although there are ways around it (like building a composite of HMI images that are chosen to be very close in time to each scanning step of an SP map) this was not done in this study. The large differences in spatial and spectral resolution are probably the biggest caveat for an inter-instrument comparison. The alignment process (rotation and pixel scaling) required an interpolation of the HMI data, which also introduces some degree of smearing. Residual misalignments remain because the variable scanning step size of the SP --and other second order effects-- were not considered

All in all, the comparison is very promising and it suggests that HMI, given its limitations, produces comparable results to SP. A more detailed and accurate comparison is being carried out in order to understand the different sources of the disagreements. 
The bottom-line is that, for any given ground or space-based observation obtained with any instrument on any part of the solar disk, there will always be reliable photospheric vector-magnetograms at a high cadence and a consistent 1\arcsec\ spatial resolution available.
HMI data are useful on their own for studies of the evolution of photospheric magnetic fields and helioseismology, but can also fill up the gaps and/or complement observations from other instruments

\acknowledgements The National Center for Atmospheric Research is partly sponsored by the National Science Foundation.

\bibliography{centeno-spw6}

\end{document}